\documentclass[aps,pre,twocolumn,10pt,amsmath,amsfonts,amssymb,floatfix,bibnotes]{revtex4-1}
\usepackage{amssymb}
\usepackage{graphicx,color}
\usepackage{bbm}
\usepackage{multirow}

\usepackage{amsmath,bm,epsfig}
\newcommand{\B}[1]{{\bm{#1}}}
\newcommand{\C}[1]{{\mathcal{#1}}}    
\DeclareMathOperator\erfc{erfc}

\begin{document}

\title{Force Distributions in Frictional Granular Media}

\author{V.S. Akella$^1$}
\author{M. M. Bandi$^1$}
\author{H. George E. Hentschel$^2$}
\author{ Itamar Procaccia$^{2}$}
\author{Saikat Roy$^2$}
\affiliation{$^1$ Collective Interactions Unit, OIST Graduate University, Onna, Okinawa, 904-0495 Japan. \\
$^2$Dept of Chemical Physics, The Weizmann Institute of Science, Rehovot 76100, Israel.}

\begin{abstract}
We report a joint experimental and theoretical investigation of the probability distribution
functions (pdf's) of the normal and tangential (frictional) forces in amorphous frictional media. We consider both the joint pdf of normal and tangential forces together, and the marginal pdf's of normal forces
separately and tangential forces separately.  A maximum entropy formalism is utilized for all these cases after identifying the appropriate constraints. Excellent agreements with both experimental and simulational data are reported. The proposed joint pdf (which appears new to the literature) predicts giant slip events at low pressures, again in agreement with observations.
\end{abstract}
\maketitle

\section{Introduction}

\begin{figure}[h!]
\includegraphics[scale=0.08]{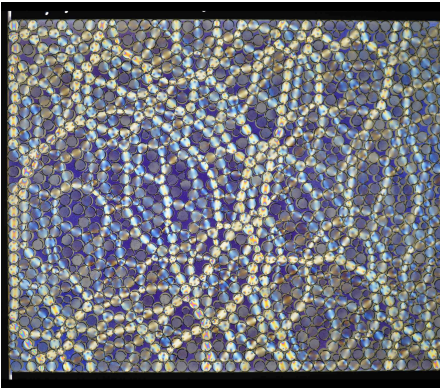}
\caption{A typical visualisation of the force chains that hold together a compressed assembly of frictional granular photoelastic disks in 2 dimensions.}
\label{forcechains}
\end{figure}

In compressed frictional amorphous granular media the external pressure is balanced by normal and tangential (frictional) forces acting
at the contacts between the grains \cite{16GPPSZ}
. The forces are
very inhomogeneous, with a wide distribution of magnitude, resulting in the appearance of force-chains
which represent the largest forces which are percolating from wall to wall, see Fig.~\ref{forcechains}.

Intensive discussions of the nature of these inter-particle forces and of their distributions in frictional amorphous media have been taking place for a number of decades. In 1995 Radjai and Roux \cite{95RR} proposed that the probability distribution function (pdf) $P_1(F^{(n)})$ of the normal contact forces $F^{(n)}_{ij}$ between grains $i$ and $j$ has a different form for forces smaller or larger than the mean normal force $\langle F^{(n)} \rangle$:
\begin{eqnarray}
P_1(F^{(n)}) &\propto& \left (\frac{F^{(n)}}{\langle F^{(n)} \rangle}\right)^\alpha \ , \quad F^{(n)}< \langle F^{(n)} \rangle \ , \\
P_1(F^{(n)}) &\propto& \exp\left[\beta\left(1-\frac{F^{(n)}}{\langle F^{(n)} \rangle}\right)\right]  , \quad F^{(n)}> \langle F^{(n)} \rangle \ .
\end{eqnarray}
A similar expression was proposed for the pdf of the tangential frictional forces $F^{(t)}_{ij}$.
A different expression was offered in the same year by the Chicago group \cite{95LNSCMNW}. This expression followed a theoretical model with the result
\begin{equation}
P_1(F^{(n)})=\frac{k^k}{(k-1)!}\left (\frac{F^{(n)}}{\langle F^{(n)} \rangle}\right)^{k-1}
\exp{\left[-k\left (\frac{F^{(n)}}{\langle F^{(n)} \rangle}\right)\right]} \ .
\label{qmodel}
\end{equation}
As discussed by Thornton \cite{97Tho}, this expression changes from an exponential distribution to
an almost Gaussian distribution as the parameter $k$ is varied from 1 to 12.

One year later, in 1996, Miller, O'Hern and Behringer concluded on the basis of careful measurements that the model
leading to Eq.(\ref{qmodel}) may miss important correlation effects leading to disagreements with
Eq.~(\ref{qmodel}) \cite{96MOB}. In other words, these measurement indicated that a relevant pdf that needs to be studied is the joint pdf $P_2(F^{(n)},F^{(t)})$. In fact, not much is reported in the literature about the effects of correlations between the normal and tangential forces. One of the aims of this paper is to close this gap.

 A few years later, in 2000, Antony \cite{00Ant} noted that for values
smaller than the average, the pdf of the normal forces can be fit with a ``half Gaussian distribution"
having 4 free parameters. For forces larger than the average the pdf was declared to be exponential.
At larger values of the strain the pdf for forces smaller than the averages was found by Anthony
to conform with a polynomial fit. One year later, in 2001, Blair et al \cite{01BMMJN} found force
distributions that ``were well represented in all cases by the functional form":
\begin{equation}
P_1(F^{(n)}) =a \left(1-b ~\exp\left[-c \left(F^{(n)}\right)^2\right] \right) \exp\left[-d F^{(n)}\right] \ ,
\end{equation}
with $a, b, c$ and $d$ being free parameters.
Yet a few years later, in 2005, Corwin, Jaeger and Nagel \cite{05CJN} offered a prediction that for Herzian contacts the pdf of the normal force should read
\begin{equation}
P_1(F^{(n)})=\alpha\left[1+\left(F^{(n)}\right)^{2/3} \frac{\langle \Delta \rangle}{d}\right]^2
\exp\left[\frac{-\beta\left(F^{(n)}\right)^{5/3}}{\beta_0} \right] \ ,
\end{equation}
where $\langle \Delta \rangle$ is the average deformation of the granules.
In the same year Majmudar and Behringer published their seminal paper in which they showed
how to visualize the forces in frictional granular matter by using photo-elastic disks \cite{05MB}.
They could show that the distributions of both the normal and the tangential forces (normalized by the mean normal force) depended on the type of external strain. The normal
force distribution for the sheared system had a peak around
the mean, a roughly exponential tail and a dip towards zero for forces
lower than the mean. In contrast, for isotropically compressed
systems, the normal force distribution dipped towards zero
for forces below the mean, was broad around the mean, and decayed
faster for large forces compared to the sheared system. The tangential
force distributions had a nearly exponential tail for forces larger than
the mean for both the sheared and the isotropically
compressed system.

The intervening years until the present time did not resolve the somewhat confusing
status of the pdf's of the contact forces in frictional matter. An interesting line of attempts
to nail down a solid prediction for these pdf's had employed the principle of maximum
entropy subject to known constraints \cite{80Sha,03Bag,04God,07HHC}. In some degree these
attempts were motivated by the desire to define an ``effective grain temperature". In the view
of the present authors these attempts were somewhat rigid in following the example of
statistical mechanics in trying to use the mean energy or the mean stress as the
appropriate constraint (on top of normalization) under which the entropy is maximized.
In statistical mechanics, as observed by Feynman \cite{72Fey}, the only ``legal" constraint is the
mean energy since the predictions of the theory must be invariant to a re-definition
of the zero-point energy $E_0$. The ratio of the probabilities to observe two states of
energy $E_1$ and $E_2$, i.e. $\exp[(E_1-E_2)/k_B T]$ must remain invariant to changes in the
reference point $E_0$. If we added as a constraint, say, $\langle E^2\rangle$, the exponential
would include a quadratic term that were not invariant to changes in the zero point energy.
This restriction is not relevant for the problem at hand. The forces between granules
are naturally bounded by zero from below, and we can use any moment of the force
distribution that appears appropriate. In this way we can reap the benefit of the information-theoretic meaning of the maximum entropy principle, providing us with the ``least biased"
prediction subject to measurable data \cite{64SW,79LT}. In fact we will show below that our
measurements of the marginal pdf $P_1$ in both experiments and simulations agree very well with the predictions of maximal
entropy subject to the mean and variance of the distributions. For the joint pdf $P_2$ one needs
to add the correlation function that couples the normal and the tangential forces.
In principle one could add additional moments as constraints but we found the agreement
with the data so good that this was (so far) deemed unnecessary.

To test the predictions of our approach we have measured the normal and tangential forces
in frictional granular matter in both experiments and simulations. We start the paper in Sect.~\ref{experiments} by describing the experimental details and the resulting
force measurements, together with similar measurements in numerical simulations. The next section, Sect.~\ref{maxent},
presents the maximum entropy approach; we calculate the predicted marginal pdf's of the magnitudes
of the forces, both normal and tangential. In Sect.~\ref{comparison} we
compare the theoretical predictions to the results of experiments and numerical simulations. In Sect.~\ref{slips} we turn to the joint pdf $P_2$. We explain that the correlations between normal and tangential forces become particularly important at low pressure. There the theoretical pdf's predict a giant frictional slip when a compressed frictional assembly is decompressed. Simulational evidence for the existence of this giant slip is presented as well. Finally, Sect.~\ref{conclusions} offers a summary and
some concluding remarks.

\section{Experiments and Numerical Simulations}
\label{experiments}

In this section we present results of experiments and simulations in which both tangential
and normal forces were measured. Here we focus on uniaxial straining and build the numerical
simulations to mimic the experimental set up. The reader who is mainly interested in the
resulting pdf's can jump directly to Subsect.~\ref{results} in which these are presented.

\subsection{Experimental Information}

\begin{figure}
\includegraphics[scale=0.20]{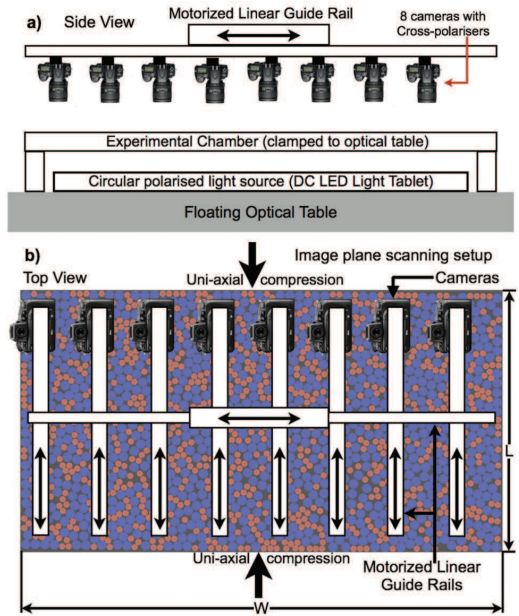}
\caption{(Color online) Experimental schematic and Image plane scanning setup (a) Side view and (b) Top view.}
\label{figmahesh}
\end{figure}

{\it Experimental Setup:} The schematic of the experimental setup is displayed in Fig.~\ref{figmahesh}. The setup was comprised of a chamber of inner dimensions 0.6 m in length, 1.1 m in width, and 0.02 m in height constructed from a steel frame with a transparent acrylic bottom plate. The chamber was lined with internal steel boundaries extending 5 cm into the chamber and connected with linear bearings that passed through the frame to rigid outer boundaries terminating in force sensors, thus setting effective inner chamber dimensions of 0.5 m length ($L$), 1 m width ($W$), and the height 0.02 m ($H$) kept unchanged. The two opposing boundaries along the $L$-axis were movable and provided uni-axial compression (see Fig.~\ref{figmahesh}a) whereas the transverse boundaries were held fixed. The chamber was rigidly clamped flat to an optical table on grade concrete flooring and floated with compressed air. A circular polarized DC light source (LED light tablet) was placed underneath the chamber to provide backlit illumination (see Fig.~\ref{figmahesh}b). The granular medium placed within the quasi-two dimensional chamber consisted of a bidispersed set of photoelastic (birefringent response to stress) disks of diameters $D_L = 1.5$ cm for large and $D_S = 1$ cm for small disks. Full details of the quasi-static translation and boundary force detection methods are presented in Ref. \cite{Bandi2017pp}.

{\it Photoelastic Disks:} The photoelastic disks were made in-house by casting liquid polymer (SQ-2001 Epoxy Resin with SQ-3154 Hardener from Avipol, Brazil) in silicone molds (Shin-Etsu Silicones, Japan) in order to control both the modulus and friction coefficient of the photoelastic disks. The silicone mold base was maintained smooth but the walls were intentionally designed to prescribed roughness to control the disk friction coefficient. The roughness was selected from industrial standard sandpaper grit chart to transfer sandpaper imprint onto silicone mold, whose imprint in turn was transferred to disk walls during polymer curing process. The disk bottom was left smooth to avoid both friction with the bottom acrylic plate as well as to permit clear transmission of light from the circular polarized backlit display. The resin-hardener mix poured into the silicone mold was baked with a free surface to allow for thermal expansion during the curing process and therefore avoid pre-stresses from developing within the disks. As a result, the resin-hardener mix poured into the moulds formed a meniscus with mold walls and cured with uneven top facet. The top faces of the cured photoelastic disks were ground on abrasive wheel to obtain clean facets with a final disk thickness of 0.975 cm.

The disk modulus was tuned by matching the epoxy resin and hardener mixture as well as the curing temperature -- the curing process being exothermic, the curing temperature for the oven had to be determined by trial-and-error to obtain disks of desired moduli. Full details of the photoelastic materials methods will be presented in a separate article, but for disks employed in the present experiments the curing temperature was set fixed at 70$^{\circ}$C for a 24 hour period and the resin-hardener mix was changed to obtain two different elastic moduli of $E = 0.004$ GPa at friction coefficient $\mu = 0.27$ and $E = 0.4$ GPa at friction coefficient of $\mu = 0.4$. Since the friction coefficient resulting from a chosen roughness on the sandpaper grit chart is not known {\it a priori}, the friction coefficient was separately measured by the method explained in Ref. \cite{Bandi2013}.

{\it Imaging:} A single digital still camera, no matter how high its resolution, does not provide the desired image quality for a quasi two-dimensional granular configuration spanning 0.5 m $\times$ 1 m. We implemented an image plane scanning system (see Fig.~\ref{figmahesh} for schematic) so it could expressly meet two design criteria. First, the large system size renders any image susceptible to angular distortions, commonly known as the fisheye effect. Whereas disks directly under the camera lens are viewed normal to the imaging plane, those farthest from the lens are at an oblique angle do not appear as circular disks but as ellipsoids instead and lead to large errors in detection of disk centers and contact stresses. Avoidance of the fisheye distortion demands moving the camera vertically higher but it drastically reduces resolution of acquired image because most of the imaged area extends outside the setup. Although disk centers are still identified by image analysis algorithms, fringe detection of photoelastic stress measurement suffers considerably. Second, the high precision quasi-static translation of 500 nm per quasi-static step achieved in this setup \cite{Bandi2017pp} demands disk displacement tracking of at least similar order. This requirement is not relevant for the current experiments as they involved a static configuration at a prescribed global pressure. Nonetheless, it becomes important for experimental analyses planned for the future.

In order to meet the above requirements, we constructed a scanning setup with eight Nikon D800E still photography cameras mounted in a row on motorized linear guide rails as shown in Fig.~\ref{figmahesh}. The eight cameras scanned the image plane providing a set of images spanning sections of the entire configuration that were digitally stitched into a composite image of size 90,000 $\times$ 180,000 pixels. The composite image had an image resolution of 1.1 $\mu$m per pixel. Although not relevant in current experiments, further improvement in image resolution from 1.1 $\mu$m to 500 nm was achieved with sub-pixel interpolation using neighboring pixel intensity values. Finally, standard granular photoelastic experiments acquire two images \cite{Daniels2017, Iikawa2016}, one without the circular cross-polarizer mounted on camera lens for disk center detection and a second image with the cross-polarizer on for photoelastic fringe detection. Our setup acquires a single image with the circular cross-polarizer on and both the disk center and photoelastic fringe detection are implemented in post-processing analysis of acquired images in two separate passes as explained below.

{\it Image Analysis:} The acquired composite image of the pack configuration was processed in two stages. Owing to backlit illumination, each disk has an illuminated ring along its edge due to diffraction bending of light with sharp intensity gradient relative to photoelastic fringe signals which possess more gradual intensity gradients. In the first stage of image processing, we applied a High-pass Gaussian convolution filter thresholded against an intensity wavenumber (inverse of distance over which the diffraction-induced intensity gradient acts). Upon applying this filter, all wavenumbers higher than the threshold wavenumber are retained in the image and all wavenumbers below it are removed. Ergo, the high-pass Gaussian convolution permits one to treat the disk edge diffraction-induced intensity as signal and photoelastic fringe intensity as noise in the first stage. We then applied a multiplicative variant of standard (additive) circular Hough transform \cite{Bandi2013}. Knowledge of total number of large and small disks and their respective radii in pixel units readily permits accurate detection of all disk centers.

In the second stage, we subtracted the High-pass Gaussian convoluted image of first stage from the original image. The resultant image now retains only photoelastic fringe intensities which were then processed using the open source Photo-elastic grain solver (PEGS) algorithms \cite{Daniels2017, PEGS} to obtain the normal and tangential forces at each stressed contact \cite{05MB,Majmudar2006}.

{\it Experimental Protocol:} A total of ten data sets were collected for a given modulus $E$ and friction coefficient $\mu$. Each of the ten data sets represented a different initially prepared granular configuration. For each of those configurations, the system was quasi-statically compressed in 500 nm steps and decompressed over 49 consecutive cycles. In the 50th compression cycle, the quasi-static compression was stopped once the boundary force sensors registered a boundary pressure value chosen {\it a priori}. For the experimental runs with disk material modulus $E = 0.004$ GPa and friction coefficient $\mu = 0.27$, the two-dimensional boundary pressure was chosen at ${\cal P} =  20$ N/m. For a second data set with disk material modulus $ E = 0.1$ GPa and friction coefficient $\mu = 0.4$, the static two-dimensional global pressure was set at ${\cal P} = 76$ N/m for measurements. We note that the disks, especially ones with higher modulus had a photoelastic threshold below which force values could not be reliably determined. Accordingly, our pdf does not include data on forces smaller than this
 threshold. This is a limitation of the experiments which needs to be taken into account when
 comparisons with theory are presented. The experimentally measured pdf's are displayed in Subsect.~\ref{results}.

\subsection{Numerical Simulations}
\label{simulations}

{\em Frictionless} granular materials are commonly studied in quasi-static protocols involving conjugate gradient methods to bring the system to mechanical equilibrium after every straining step \cite{97MLGLBW}; but when the particles have friction, Molecular Dynamics simulations are preferred as they correctly keep track of both the normal and the (history dependent) tangential forces \cite{16GPP}. So we set up simulation of uniaxial compression of two dimensional granular packings, performed using open source codes, LAMMPS \cite{95P} and LIGGGHTS \cite{12KGHAP}. To mimick the experimental system the particles are taken as bi-dispersed disks of unit mass with diameters 1 and $1.4$ respectively. All the lengths in the simulations are measured in units of the small diameter. The particles are placed randomly in a three dimensional box of dimension, $57$ (along $x$), $102$ (along $y$) and $1.4$ (along $z$). Quasistatic compression is implemented by displacing the boundary particles. A side wall made of particles is placed in the direction perpendicular to the compression direction.

The contact forces (both the normal and tangential forces which arise due to friction) are modeled according to the discrete element method developed by Cundall and Strack \cite{79CS}.
When the disks are compressed they interact via both
normal and tangential forces.  Particles $i$ and $j$, at positions ${\B r_i, \B r_j}$ with velocities ${\B v_i, \B v_j}$ and angular velocities ${\B \omega_i, \B \omega_j}$ will  experience a relative normal compression on contact given by $\Delta_{ij}=|\B r_{ij}-D_{ij}|$, where $\B r_{ij}$ is the vector joining the centers of mass and $D_{ij}=R_i+R_j$; this gives rise to a  normal force $ \B F^{(n)}_{ij} $. The normal force is modeled as a Hertzian contact, whereas the tangential force is given by a Mindlin force \cite{79CS}. Defining $R_{ij}^{-1}\equiv R_i^{-1}+R_j^{-1}$, the force magnitudes are,
\begin{eqnarray}
\B F^{(n)}_{ij}&=&k_n\Delta_{ij} \B n_{ij}-\frac{\gamma_n}{2} \B {v}_{n_{ij}}\ , \:
\B F^{(t)}_{ij}=-k_t \B t_{ij}-\frac{\gamma_t}{2} \B {v}_{t_{ij}} \\
k_n &=& k_n^{'}\sqrt{ \Delta_{ij} R_{ij}} \ , \quad
k_t = k_t^{'} \sqrt{ \Delta_{ij} R_{ij}} \\
\gamma_{n} &=& \gamma_{n}^{'}  \sqrt{ \Delta_{ij} R_{ij}}\ , \quad
\gamma_{t} = \gamma_{t}^{'}  \sqrt{ \Delta_{ij} R_{ij}} \ .
\end{eqnarray}
Here $\delta _{ij}$ and $t_{ij}$ are normal and tangential displacement; $r_{ij}$  is the effective radius. $\B n_{ij}$ is the normal unit vector.  $k_n^{'}$ and $k_t^{'}$ are spring stiffness for normal and tangential mode of deformation: $\gamma_n^{'}$ and $\gamma_t^{'}$ are viscoelastic damping constant for normal and tangential deformation.
   $\B {v_n}_{ij}$ and $\B {v_t}_{ij}$ are respectively normal and tangential component of the relative velocity between two particles. The relative normal and tangential velocity are given by
   \begin{eqnarray}
\B {v}_{n_{ij}}&=& (\B {v}_{ij} .\B n_{ij})\B n_{ij}  \\
\B {v}_{t_{ij}}&=& \B {v}_{ij}-\B {v}_{n_{ij}} - \frac{1}{2}(\B \omega_i + \B \omega_j)\times \B r_{ij}.
\end{eqnarray}
   where $\B {v}_{ij} = \B {v}_{i} - \B {v}_{j}$. Elastic tangential displacement $ \B t_{ij}$ is set to zero when the contact is first made and is calculated using $\frac{d \B t_{ij}}{d t}= \B {v}_{t_{ij}}$ and also the rigid body rotation around the contact point is accounted for to ensure that $ \B t_{ij}$ always remains in the local tangent plane of the contact \cite{01SEGHLP}.

   The translational and rotational acceleration of particles are calculated from Newton's second law; total forces and torques on particle $i$ are given by

      \begin{eqnarray}
\B F^{(tot)}_{i}&=& \sum_{j}\B F^{(n)}_{ij} + \B F^{(t)}_{ij}  \\
\B \tau ^{(tot)}_{i}&=& -\frac{1}{2}\sum_{j}\B r^{ij} \times \B F^{(t)}_{ij}.
\end{eqnarray}

   The tangential force varies linearly with the relative tangential displacement at the contact point as long as the tangential
   force does not exceed the limit set by the Coulomb limit
   \begin{equation}
   F^{(t)}_{ij} \le \mu F^{(n)}_{ij} \ , \label{Coulomb}
   \end{equation}
  where $\mu$ is a material dependent coefficient. When this limit is exceeded the contact slips in a dissipative
  fashion. In our simulations we reset the
  value of $t_{ij}$  so that $F^{(t)}_{ij} =0.8 \mu F^{(n)}_{ij}$. This choice is somewhat arbitrary, but recommended on the basis of frictional slip events measured in
  experiments in the laboratory of J. Fineberg \cite{Fine}. A global damping is implemented to reach the static equilibrium in reasonable amount of time. After each compression step, a relaxation step is added so that the system reaches the static equilibrium and then the forces at all the contacts are measured. In addition the global stress tensor is measured by taking averages of the dyadic products between the contact forces and the branch vector over all the contacts in a given volume,
  \begin{equation}
\sigma_{\alpha \beta} =\frac{1}{V}\sum_{j\neq i}\frac{r^{\alpha}_{ij} F^{\alpha}_{ij} }{2}
  \end{equation}
The pressure $\C P$ is determined from the trace of the stress. The resulting distribution of forces
are presented in the next subsection.

\subsection{The resulting normal and tangential force distributions}
\label{results}

\begin{figure}
\includegraphics[scale=0.65]{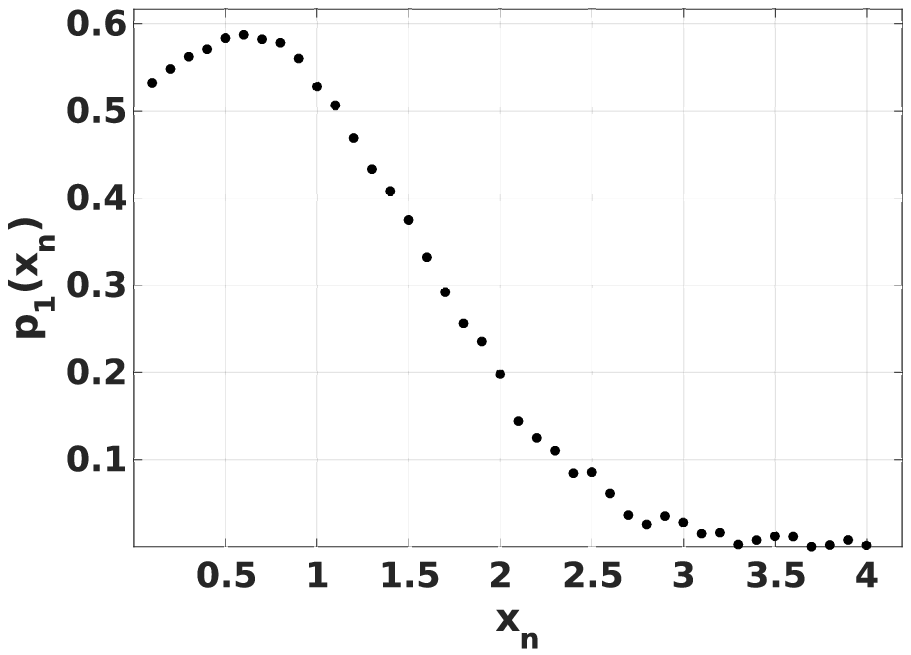}
\includegraphics[scale=0.65]{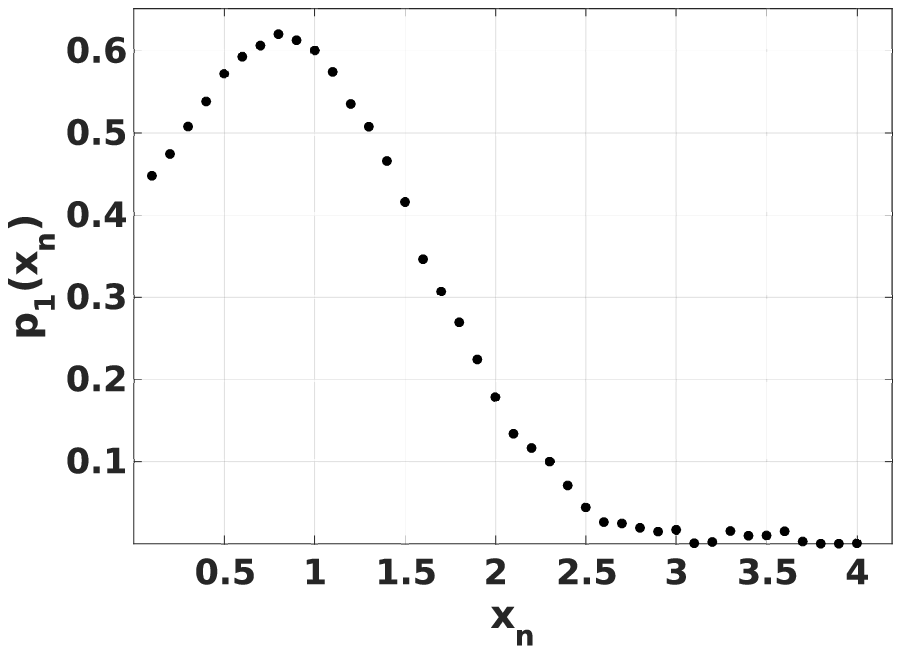}
\caption{The pdf's of the mean-normalized normal forces as measured in the
experiment. Upper panel: $\mu=0.4$ and $\C P=76$ N/m. Lower panel $\mu=0.27$ and $\C P=20$ N/m. }
\label{exp1}
\end{figure}
\begin{figure}
\includegraphics[scale=0.65]{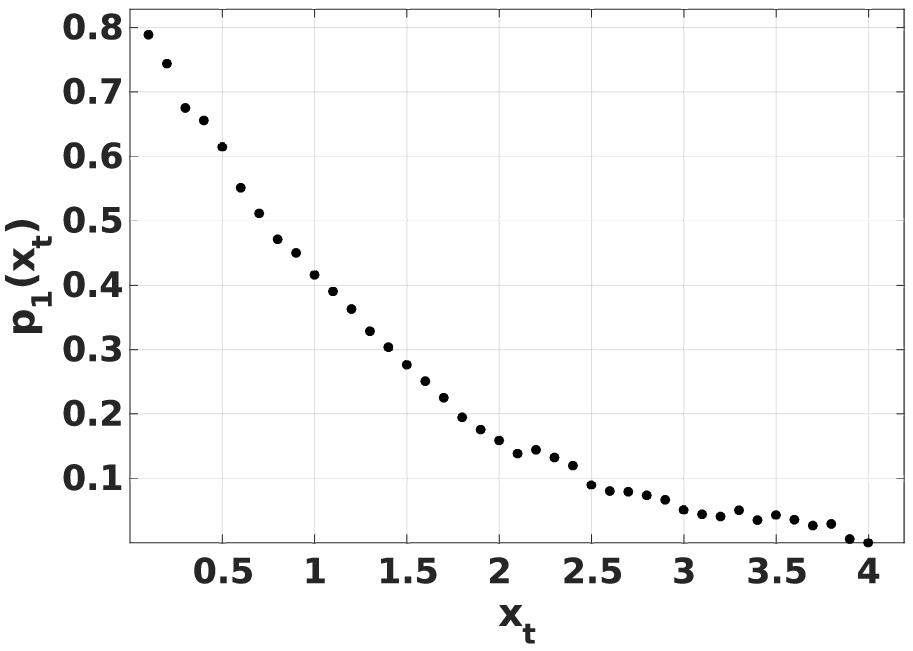}
\includegraphics[scale=0.65]{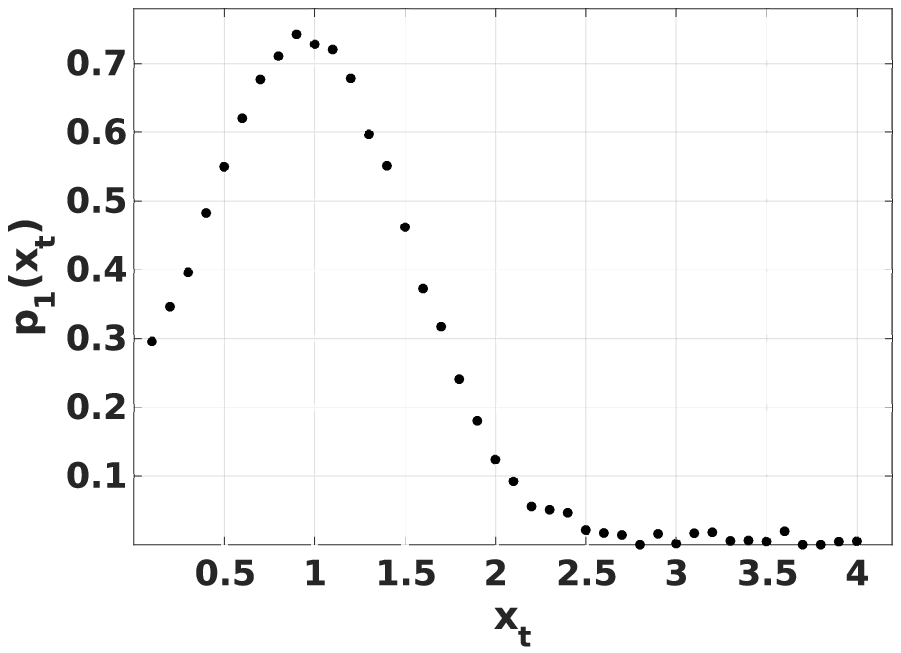}
\caption{The pdf's of the mean-normalized tangential forces as measured in the
experiment. Upper panel: $\mu=0.4$ and $\C P=76$ N/m. Lower panel $\mu=0.27$ and $\C P=20$ N/m. }
\label{exp2}
\end{figure}

In discussing the force distributions one can consider in principle a number of different pdf's
of varying complexity.
As said in the introduction, the forces acting on the contacts of grains in frictional amorphous
matter are highly inhomogeneous. Thus for $N$ particles there exists a complex joint probability distribution $P_N(\{F^{(n)}\}, \{F^{(t)})\}$ for the magnitudes of these forces, where we have used  the notation
\begin{equation}
\{F^{(n)}\}\equiv \{F^{(n)}_{ij}; i,j \text{ running on all contacts}\} \ ,
 \end{equation}
 for the normal forces, and similarly for the tangential forces. Integrating over all contacts except those for one pair of connected particles we can define the joint probability distribution $P_2(F^{(n)}, F^{(t)})$; while for the normal and transverse forces separately we can define the probability distribution for the normal forces:
\begin{equation}
P_1(F^{(n)})=\int_0^\infty P_2(F^{(n)},F^{(t)})  d F^{(t)} \ ,
 \end{equation}
For the transverse forces
\begin{equation}
P_1(F^{(t)})=\int_0^\infty P_2(F^{(n)},F^{(t)})  dF^{(n)} .
 \end{equation}
 In general  $P_2(F^{(n)},F^{(t)}) \ne P_1(F^{(n)})P_1(F^{(t)})$. We will consider
first the ``single-particle" pdf $P_1$ for the normal and tangential forces. Later in Sect.~\ref{slips} we will discuss also the joint pdf $P_2$.

\subsubsection{Experimental results}
In a number of experimental and simulational studies it was found that the probabilities
$P_1(F^{(n)})$ and $P_1(F^{(t)})$ collapse nicely when plotted with the argument normalized by its mean. Accordingly we define
\begin{equation}
p_1(x_n)\equiv P_1\left(\frac{F^{(n)}}{\langle F^{(n)}\rangle} \right) \ , \quad
p_1(x_t)\equiv P_1\left(\frac{F^{(t)}}{\langle F^{(t)}\rangle} \right) \ .
\end{equation}

Here we present the pdf's $p_1(x_n)$ and $p_1(x_t)$ which were measured as explained in the experimental protocol above. In Fig.~\ref{exp1} find the pdf's of the normal forces at two different pressures $\C P=76$ N/m and $\C P=20$ N/m.
The corresponding pdf's for the tangential mean-normalized forces are presented in Figs.~\ref{exp2}.

It is interesting to note that the nature of the pdf's of the tangential forces are more sensitive
to the change in parameters. The maximum which exists at both pressures for the pdf of the normal
forces and for the tangential forces at low pressures is absent in the case of the tangential
forces at high pressure.

\subsubsection{Simulation results}

\begin{figure}
\includegraphics[scale=0.65]{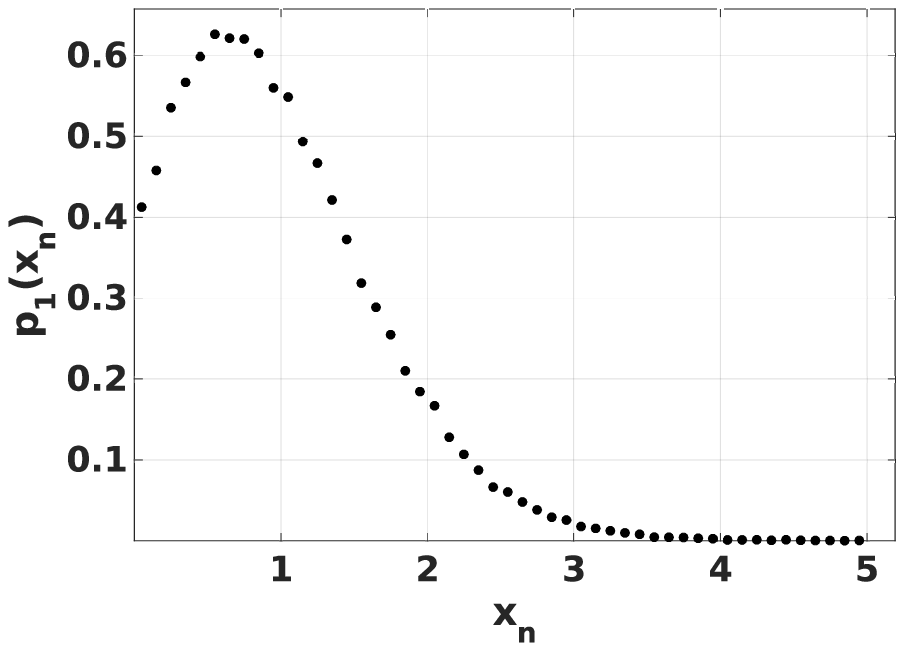}
\includegraphics[scale=0.65]{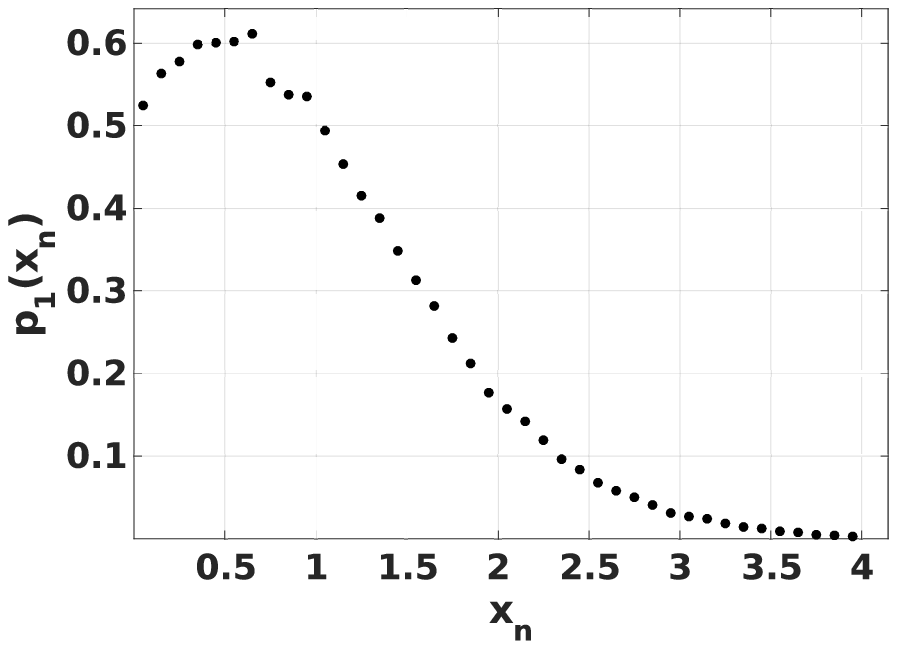}
\caption{The pdf's of the mean-normalized normal forces as measured in the
simulations. Upper panel: $\mu=0.1$ and $\C P=83.5$. Lower panel: $\mu=0.1$ and $\C P=20$. }
\label{sim1}
\end{figure}

\begin{figure}
\includegraphics[scale=0.65]{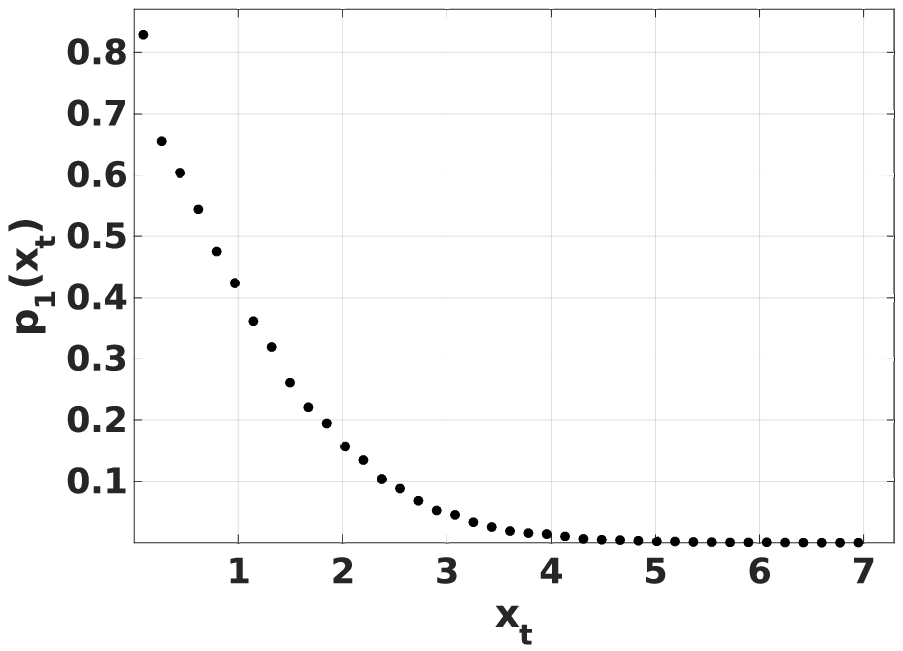}
\includegraphics[scale=0.65]{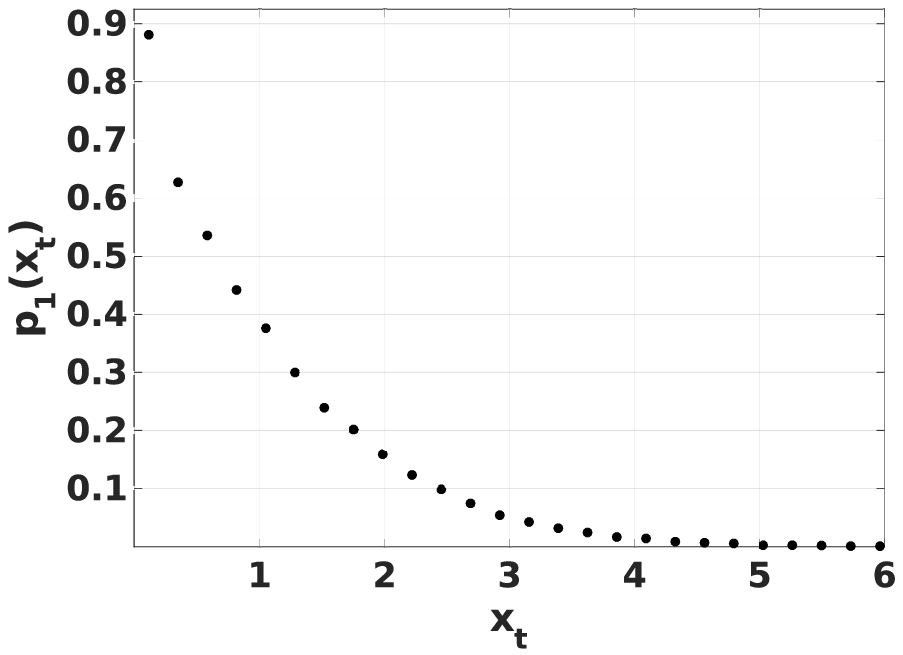}
\caption{The pdf's of the mean-normalized tangential forces as measured in the
simulations. Upper panel: $\mu=0.1$ and $\C P=83.5$. Lower panel: $\mu=0.1$ and $\C P=20$. }
\label{sim2}
\end{figure}

Here we present the  pdf's of the normal and tangential mean-normalized forces which were measured as explained in the simulation subsection above. In Fig.~\ref{sim1} we present the pdf's of the mean-normalized normal forces, again for two different values of the pressure.
The corresponding figures for the tangential mean-normalized forces are shown in
Fig.~\ref{sim2}.
We note that in the simulation results the pdf's of the tangential forces lack a maximum
for both pressures.

\section{Maximum Entropy and the Marginal pdf's of force magnitudes in frictional matter}
\label{maxent}

\begin{figure}
\includegraphics[scale=0.70]{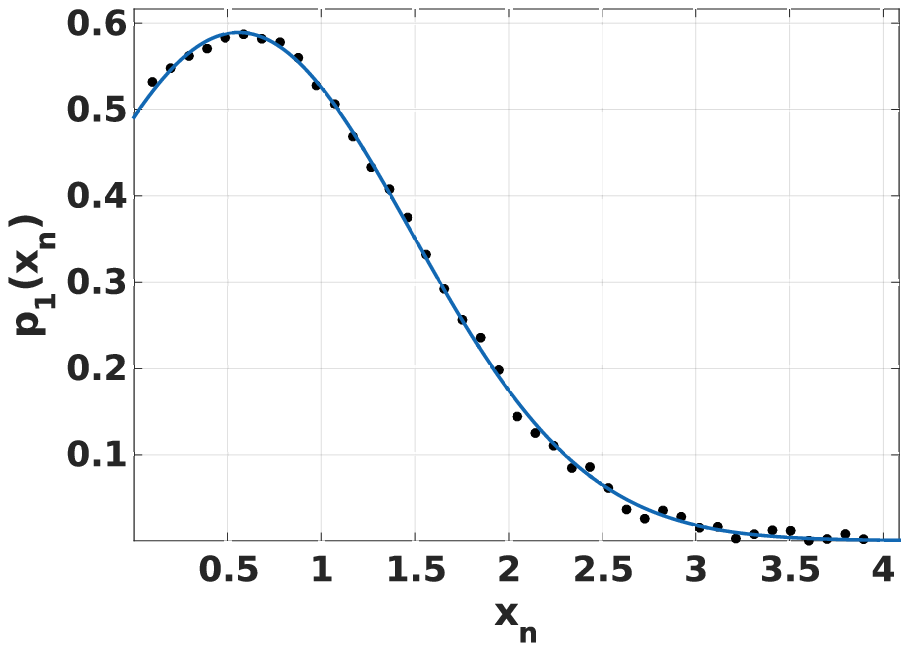}
\includegraphics[scale=0.70]{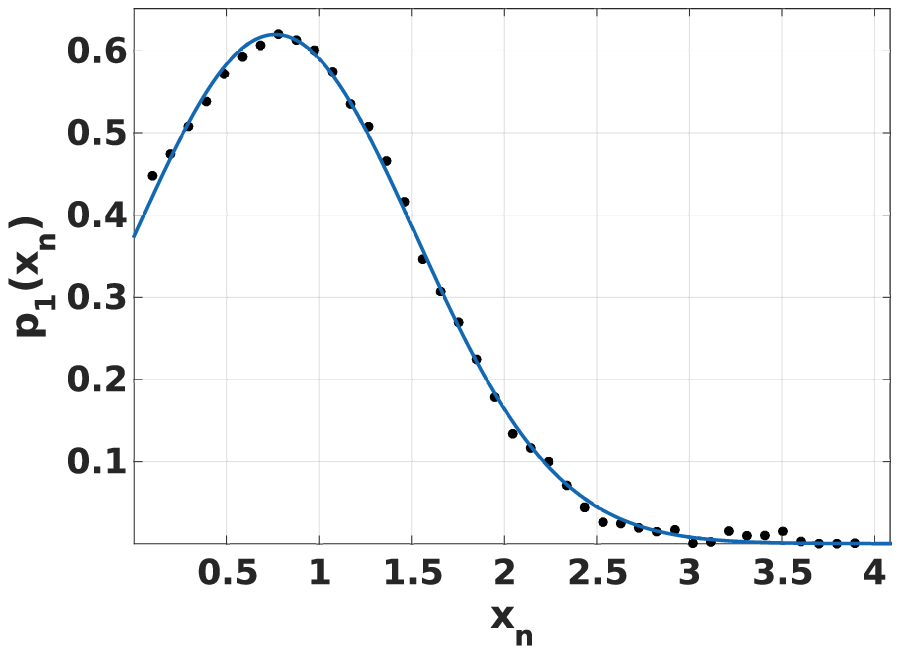}
\caption{Comparison of the functional prediction Eqs.~(\ref{Pnew}) to the pdf's of the mean-normalized normal forces as measured in the
experiments. The upper and lower panel correspond to Fig.~\ref{exp1}. In the upper panel
$\lambda_n=-0.65$ and $\lambda_{nn}=0.58$. In the lower panel $\lambda_n=-1.32$ and $\lambda_{nn}=0.87$.}
\label{comexp1}
\end{figure}
\begin{figure}
\includegraphics[scale=0.70]{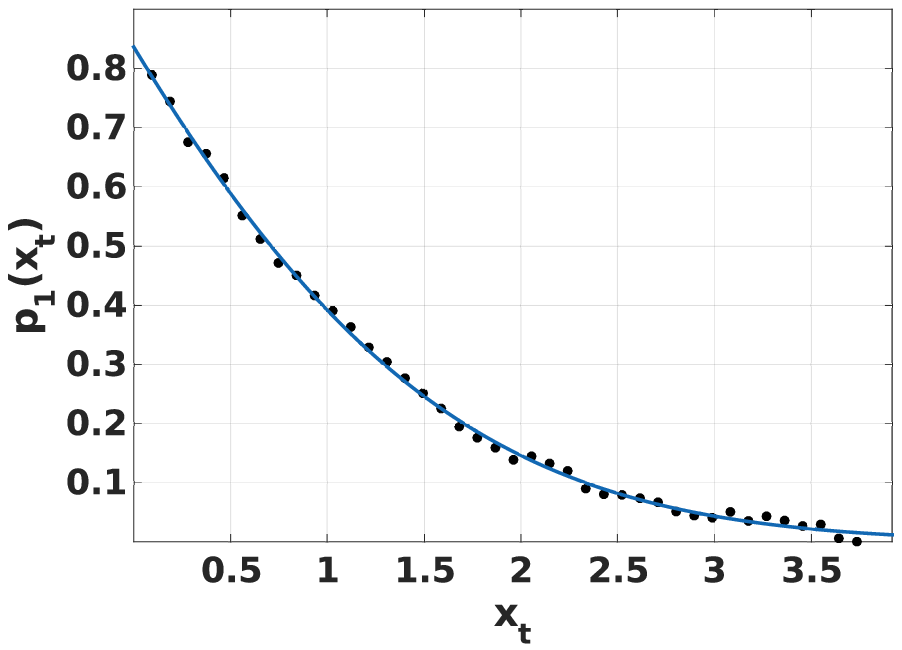}
\includegraphics[scale=0.70]{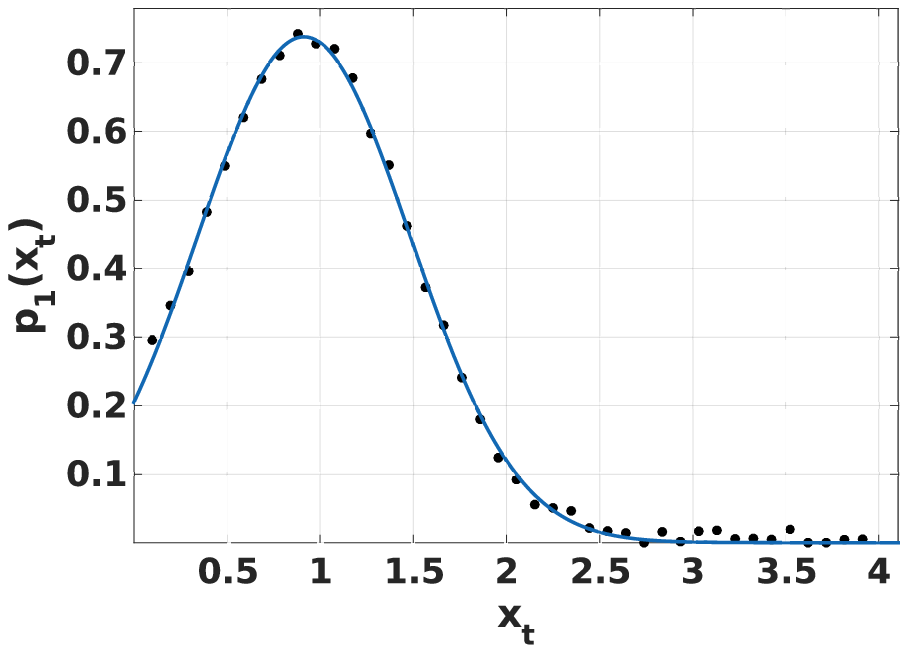}
\caption{Comparison of the functional prediction Eqs.~(\ref{Pnew}) to the pdf's of the mean-normalized tangential forces as measured in the
experiments. The upper and lower panel correspond to Fig.~\ref{exp2}. In the upper panel
$\lambda_t=0.64$ and $\lambda_{tt}=0.11$. In the lower panel $\lambda_t=-2.81$ and $\lambda_{tt}=1.54$ .}
\label{comexp2}
\end{figure}
We seek an analytic form for these pdf's by maximizing the entropy
\begin{equation}
S \equiv -\int_0^\infty p_1(x) \ln p_1(x) dx \ ,
\end{equation}
subject to constraints. Using a single constraint, i.e that $\langle x\rangle =1$ and normalizing
the pdf in the range $[0,\infty]$ yields an exponential form for the distribution
\begin{equation}
\label{Pexp}
p_1(x) =\lambda_a\exp{(-\lambda_a x )} \ , \text{(unacceptable)} \ ,
\end{equation}
for both the mean-normalized normal and tangential forces $x_n$ and $x_t$.
A glance at experimental data for the distributions of the mean-normalized normal and transverse forces in Figs.~\ref{exp1}-\ref{sim2} shows that they are not exponential as Eq.~(\ref{Pexp}) suggests.
The existence of a clear maximum in the distributions indicates that a minimal additional constraint should be provided by the variance $\sigma^2 = \langle x^2 \rangle -\langle x\rangle^2$ in both cases. Using now the mean {\em and} the variance constraints the maximum entropy formalism yields for both the normal and transverse forces similar forms for $p_1(x)$,
\begin{equation}
\label{P}
p_1(x) = \frac{ \exp{(-\lambda_a x - \lambda_b x^2)}}{Z(\lambda_a,  \lambda_b)}
\end{equation}
with the partition function
\begin{equation}
\label{Z}
 Z(\lambda_a,  \lambda_b) = \sqrt{ \frac{\pi }{4\lambda_b}} e^{\frac{\lambda_a^2}{4  \lambda_b}} \erfc\left(\frac{\lambda_a}{2\sqrt{ \lambda_b}}\right) \ .
\end{equation}
The Lagrange multipliers can be found from the partial derivatives
\begin{eqnarray}
\label{DlnZ}
 -\frac{\partial \log{Z(\lambda_a,  \lambda_b)}}{ \partial \lambda_a}
& = & 1 \ , \nonumber\\
 -\frac{\partial \log{Z(\lambda_a,  \lambda_b)}}{\partial \lambda_b}
& = & 1+ \sigma^2 \ .
\end{eqnarray}
To compute the Lagrange multipliers which are required to get explicit forms for the probability distributions let us define the associated functions $y= \lambda_a/(2 \sqrt{ \lambda_b}) $. Then from Eqs.~(\ref{DlnZ}) we derive an equation for $y$ as the nonlinear root of the equation
\begin{equation}
\label{M}
\frac{[ y^2 + \case{1}{2} - e^{-y^2} \frac{y}{\sqrt{\pi} \erfc{y}}]}{ [ -y + \frac{e^{-y^2}}{\sqrt{\pi} \erfc{y}}]^2}= 1+\sigma^2 \ .
\end{equation}
Once we solve this last equation for $y(\sigma)$ we can find the two Lagrange multipliers that fix $p_1(x)$ as
\begin{eqnarray}
\label{Eq1}
\lambda_b(\sigma) &= &[ -y(\sigma) + \frac{e^{-y^2(\sigma)}}{\sqrt{\pi} \erfc{y(\sigma)}}]^2\nonumber \\
\lambda_a(\sigma)& = &2 y(\sigma) \sqrt{\lambda_b(\sigma)}
 \end{eqnarray}
Finally  we can now write down the explicit distributions for both transverse forces $p_1(x_t)$ and the normal forces $p_1(x_n)$  as follows
\begin{eqnarray}
\label{Pnew}
p_1(x_t) & = &\frac{ \exp{(-\lambda_t x_t - \lambda_{tt} x_t^2)}}{Z(\lambda_t,  \lambda_{tt})}\nonumber \\
p_1(x_n) & = &\frac{ \exp{(-\lambda_n x_n - \lambda_{nn} x_n^2)}}{Z(\lambda_n,  \lambda_{nn})} \ ,
\end{eqnarray}
with the partition functions
\begin{eqnarray}
\label{ZZ}
 Z(\lambda_t,\lambda_{tt}) & = &\frac{\sqrt{ \pi}}{4\lambda_{tt}}} e^{\frac{\lambda_t^2}{4  \lambda_{tt}}} \erfc{\Big(\frac{\lambda_t}{2\sqrt{ \lambda_{tt}}}\Big)\nonumber \\
 Z(\lambda_n,\lambda_{nn}) & = &\frac{\sqrt{ \pi}}{4\lambda_{nn}}} e^{\frac{\lambda_n^2}{4  \lambda_{nn}}} \erfc{\Big(\frac{\lambda_n}{2\sqrt{ \lambda_{nn}}}\Big) \ .
\end{eqnarray}

\section{Comparison of experimental and simulational results to theory}
\label{comparison}
\begin{figure}
\includegraphics[scale=0.70]{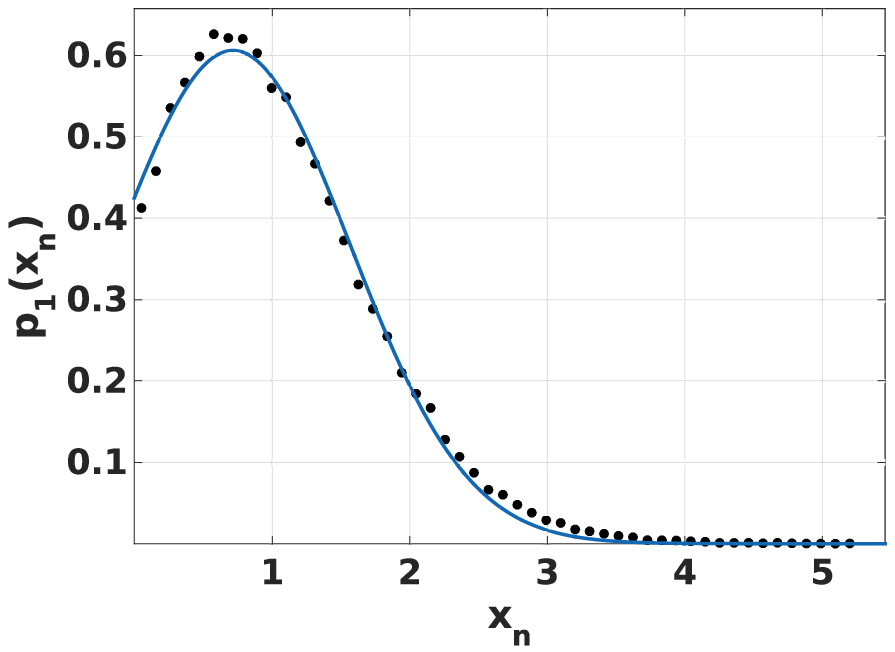}
\includegraphics[scale=0.70]{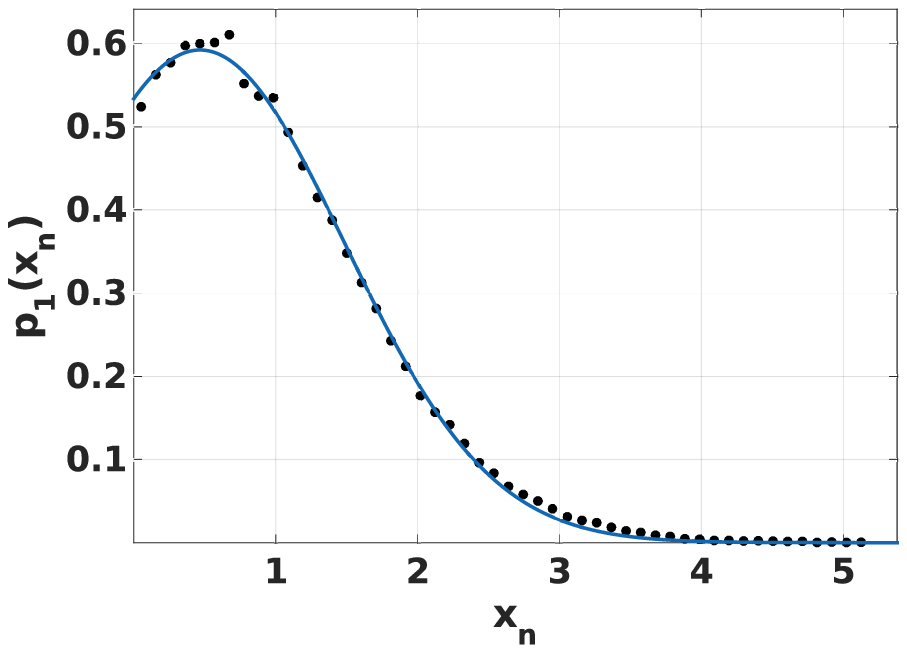}
\caption{Comparison of the functional prediction Eqs.~(\ref{Pnew}) to the pdf's of the mean-normalized normal forces as measured in the
simulations. The upper and lower panel correspond to Fig.~\ref{sim1}. In the upper
panel $\lambda_n=-0.99$, $\lambda_{nn} = 0.69$. In the lower panel $\lambda_n=-0.44$, $\lambda_{nn} = 0.48$.}
\label{comsim1}
\end{figure}

In this section we present the comparison of the theory to the measurements in experiments
and in simulations. In executing this comparison we need to be careful. The theory assumes
that we have full data for $0\le x \le \infty$ and that the normalization is computed over
the whole interval. As explained above, in the experiment we are limited in resolving the small
forces due to the optical limitation, and also very large forces suffer from lesser statistics.
In the simulations we also recognize finite size effects which limit the statistics of very
small and very large forces. Thus the measurement of the mean and variance of the pdf's
directly from the data cannot conform with the theoretical requirement of having data over
the full interval. To overcome this difficulty we have {\em fitted} the best values of
the lagrange multipliers using the data and the functional form Eq.~(\ref{Pnew}). Once we
fit the form we have a pdf over the whole interval, and we can compute the mean and the variance.
We note that the mean and variance may deviate somewhat from their counterparts which are
evaluated directly from the data. We consider the latter to be inferior since they stem
from incomplete data. We should recognize however that the definition of $x_n$ and $x_t$ involves
the average forces, and therefore in the comparison below the $x$ axes are re-scaled somewhat
differently to these axes in the pdf's shown so far. To ensure consistency,
we always check whether the theoretical values of the Lagrange multipliers are indeed in
agreement with the Eq.~(\ref{Eq1}) using the recomputed average and variance. All the results below were obtained using this
procedure and showed excellent self consistency with the theoretical numbers.

In Fig.~\ref{comexp1} we show the agreement between the theory and
the experimental measurements of the pdf's of the normal mean-normalized forces.
The corresponding comparisons for the pdf's of the mean averaged tangential forces
are shown in Fig.~\ref{comexp2}.

\begin{figure}
\includegraphics[scale=0.70]{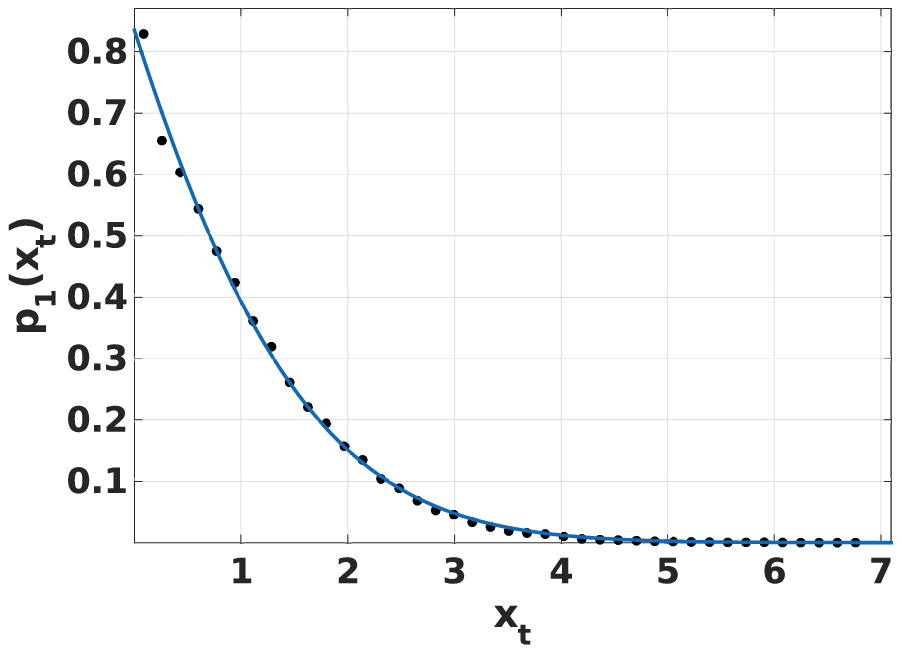}
\includegraphics[scale=0.70]{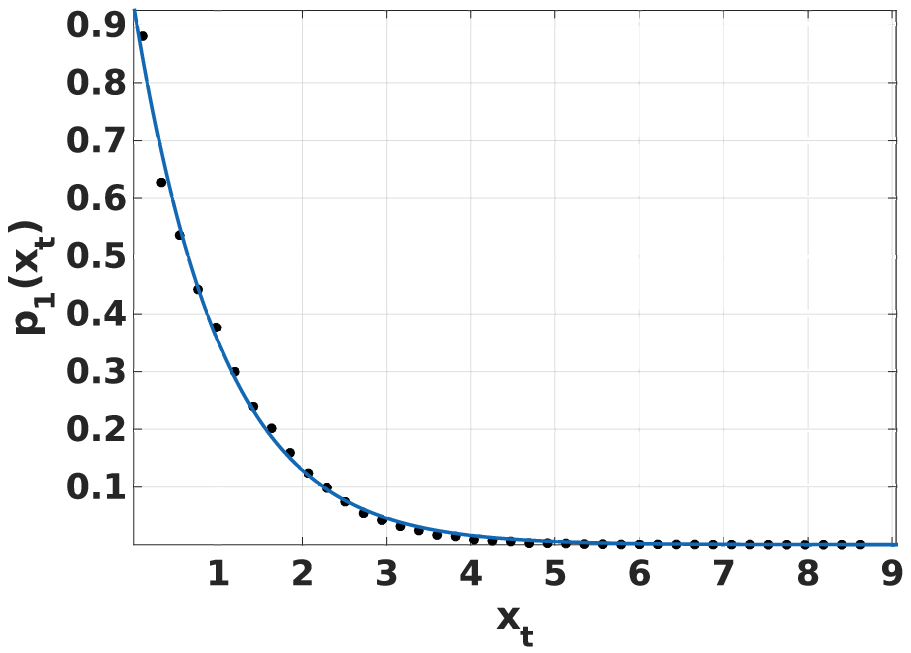}
\caption{Comparison of the functional prediction Eqs.~(\ref{Pnew}) to the pdf's of the mean-normalized tangential forces as measured in the
simulations. The upper and lower panel correspond to Fig.~\ref{sim2}. In the upper
panel $\lambda_t=0.65$, $\lambda_{tt} = 0.10$. In the lower panel $\lambda_t=0.96$, $\lambda_{tt} = 0.01$. }
\label{comsim2}
\end{figure}

The comparison of the theory to the simulations results are shown next. In Fig.~\ref{comsim1}
we present the pdf's of the mean-normalized normal forces. The upper and lower panel
correspond to Fig.~\ref{sim1}.
The corresponding comparisons for the pdf's of the mean-normalized tangential forces
from the simualations are shown in Fig.~\ref{comsim2}.
The conclusion is that at least for the data at hand, both in experiments and in simulations,
at different values of the pressure, the theoretical prediction of the analytic forms of
the $p_1(x_n)$ and $p_1(x_t)$ fit the data admirably well. We now turn to the joint probability $p_2(x_n,x_t)$ which is sensitive to the correlation between the normal and tangential forces. This will underline the predictive value of the present approach.
\section{Joint distributions and the prediction of a giant slip}
\label{slips}
In this section we study the properties of the joint distributions $p_2(x_n,x_t)$ as a function
of the pressure. As had been commented in Ref.~\cite{96MOB} correlations between the normal
and tangential forces cannot be neglected with impunity. If there were no correlations between
the normal and tangential forces, then we could expect that $\langle  x_n x_t \rangle =1$ at all pressures. Measuring this correlations in the simulations shows that this is not the case
 at {\em any} pressure.  We therefore need to take these correlations into account, expose
 the physical reason for the correlations and draw the necessary conclusions. We will argue in this
 section that the fundamental reason for the correlation is the Coulomb constraint Eq.~(\ref{Coulomb}). The most important consequence that we could find is the existence of a giant slip event
 at low pressures as described and discussed below.
\subsection{Maximum Entropy formalism for the joint distributions}
\label{jointmaxent}

 In order to employ the maximum entropy formalism for the joint distributions we need to incorporate the correlation $\langle  x_n x_t \rangle$ into the formalism.
 Measuring this correlation in either experiments or simulations and introducing an additional lagrange multiplier $\lambda_c(\C P)$ we can write down the maximum entropy joint distribution
 \begin{widetext}
\begin{equation}
\label{P2,a}
p_2(x_n,x_t) = \frac{ \exp{[-\lambda_n x_n - \lambda_{nn} x_n^2-\lambda_t x_t - \lambda_{tt} x_t^2-\lambda_c x_n x_t]}\theta(\mu\langle F^{n}\rangle x_n-\langle F^{t}\rangle x_t)}{Z_2}
\end{equation}
where the $\theta$-function is respecting the Coulomb constraint and
\begin{equation}
\label{Za}
 Z_2  =  \int_0^\infty \int_0^\infty dx_n dx_t \exp{[-\lambda_n x_n - \lambda_{nn} x_n^2-\lambda_t x_t - \lambda_{tt} x_t^2-\lambda_c x_n x_t]}\theta(\mu\langle F^{n}\rangle x_n-\langle F^{t}\rangle x_t) \ .
\end{equation}
\end{widetext}
The new Lagrange multiplier $\lambda_c$ can now be extracted from the additional equation
\begin{equation}
\label{DlnZc}
 -\partial \log{Z_2(\lambda_n,  \lambda_{nn}, \lambda_t,  \lambda_{tt},\lambda_c)}/ \partial \lambda_c = \langle x_n x_t \rangle \ .
\end{equation}

In reality it turns out that   Eqs.~(\ref{Za}) and~(\ref{DlnZc}) are somewhat difficult to invert to get an explicit expression for the five pressure-dependent lagrange multipliers $\lambda_n(\C P),\lambda_{nn}(\C P),\lambda_t(\C P),\lambda_{tt}(\C P),\lambda_c(\C P)$. For a precise calculations all these are required as the joint distribution has altered in form from our marginal expressions. At this point we are interested however in the {\em qualitative} predictions that the formalism can provide. To this aim we shall keep the four lagrange multipliers given by Eqs.~(\ref{Eq1}) for the marginal distributions, and neglect firstly all correlations between the normal and tangential forces. Then we can expand the partition function in powers of $\lambda_c$ to second order. In this approximation the normal and tangential terms become disconnected and we can write
\begin{equation}
\label{Zapprox}
Z_2 \approx Z_nZ_t[1-\lambda_c\langle x_n x_t \rangle_0 + (1/2)\lambda_c^2 \langle x_n^2x_t^2\rangle_0+\cdots \ ,
\end{equation}
where a subscript zero means the lowest order approximation of no correlation.
Now in this approximation $\langle x_nx_t \rangle_0 \approx \langle x_n\rangle\langle x_t \rangle=1$ and $\langle x_n^2x_t^2\rangle_0\approx \langle x_n^2\rangle \langle x_t^2\rangle$ and using Eq.~(\ref{DlnZc}), we find
\begin{equation}
\label{lcnew}
\langle x_n x_t \rangle = \frac{1-\lambda_c \langle x_n^2\rangle \langle x_t^2\rangle}{1- \lambda_c + (1/2) \lambda_c^2 \langle x_n^2\rangle \langle x_t^2\rangle}
\end{equation}
Eq.~(\ref{lcnew}) can be solved to get $\lambda_c$ in terms of the known second moments of the normal and tangential forces
\begin{eqnarray}
\label{lmcorr}
&&\lambda_c \approx  [1/(\langle x_n^2\rangle\rangle \langle x_t^2\rangle) - 1/\langle x_n x_t \rangle]\nonumber\\&&+\sqrt{[1/\langle x_n x_t \rangle^2-1/(\langle x_n^2\rangle\rangle \langle x_t^2\rangle)^2]}.
\end{eqnarray}

Using our simulations data we evaluated $\lambda_c(\C P)$ for any desired pressure.  We found that $\lambda_c(\C P)$ is a weak function of pressure but clearly nonzero in value.
\begin{figure}
\includegraphics[scale=0.27]{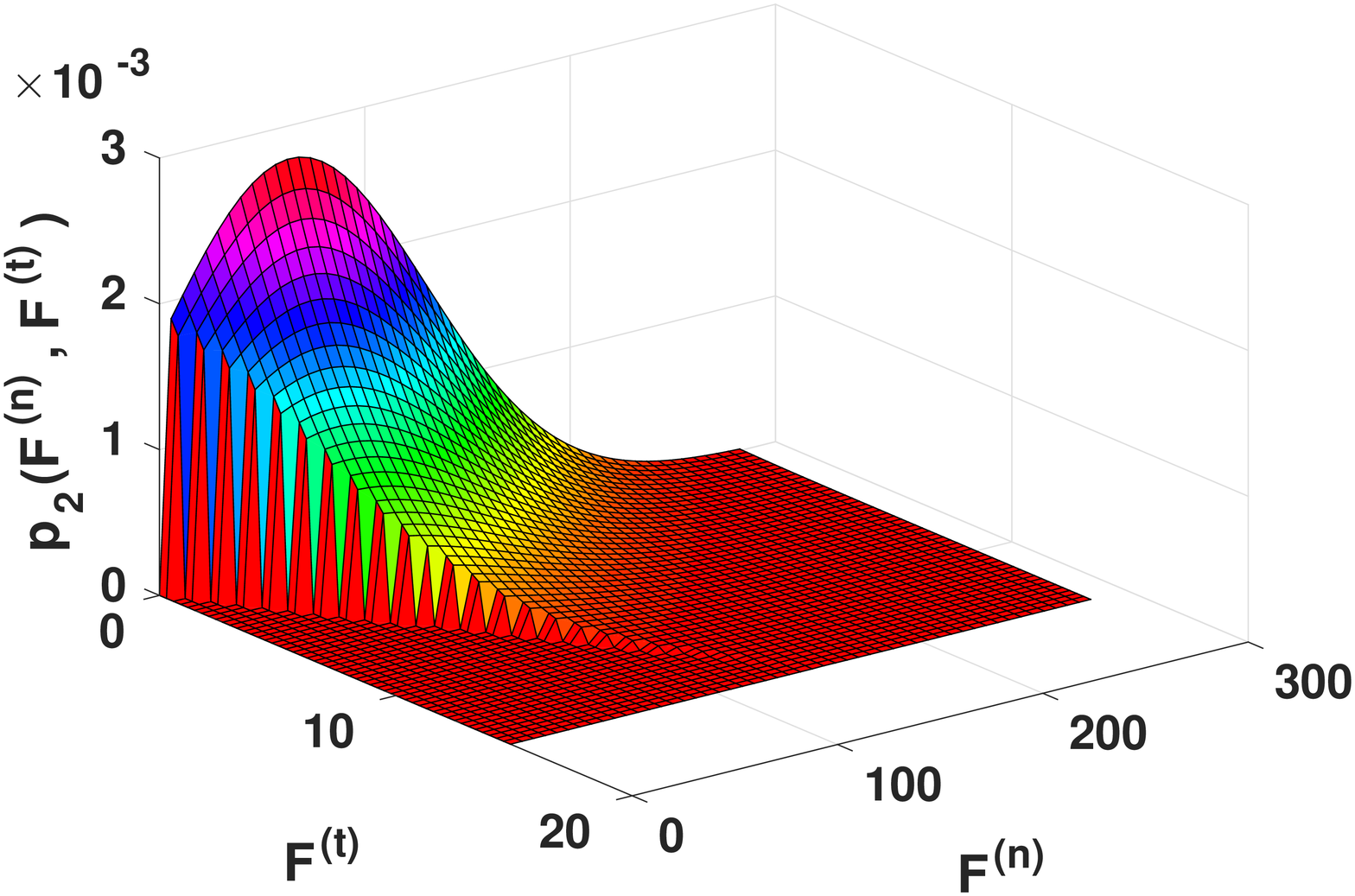}
\includegraphics[scale=0.27]{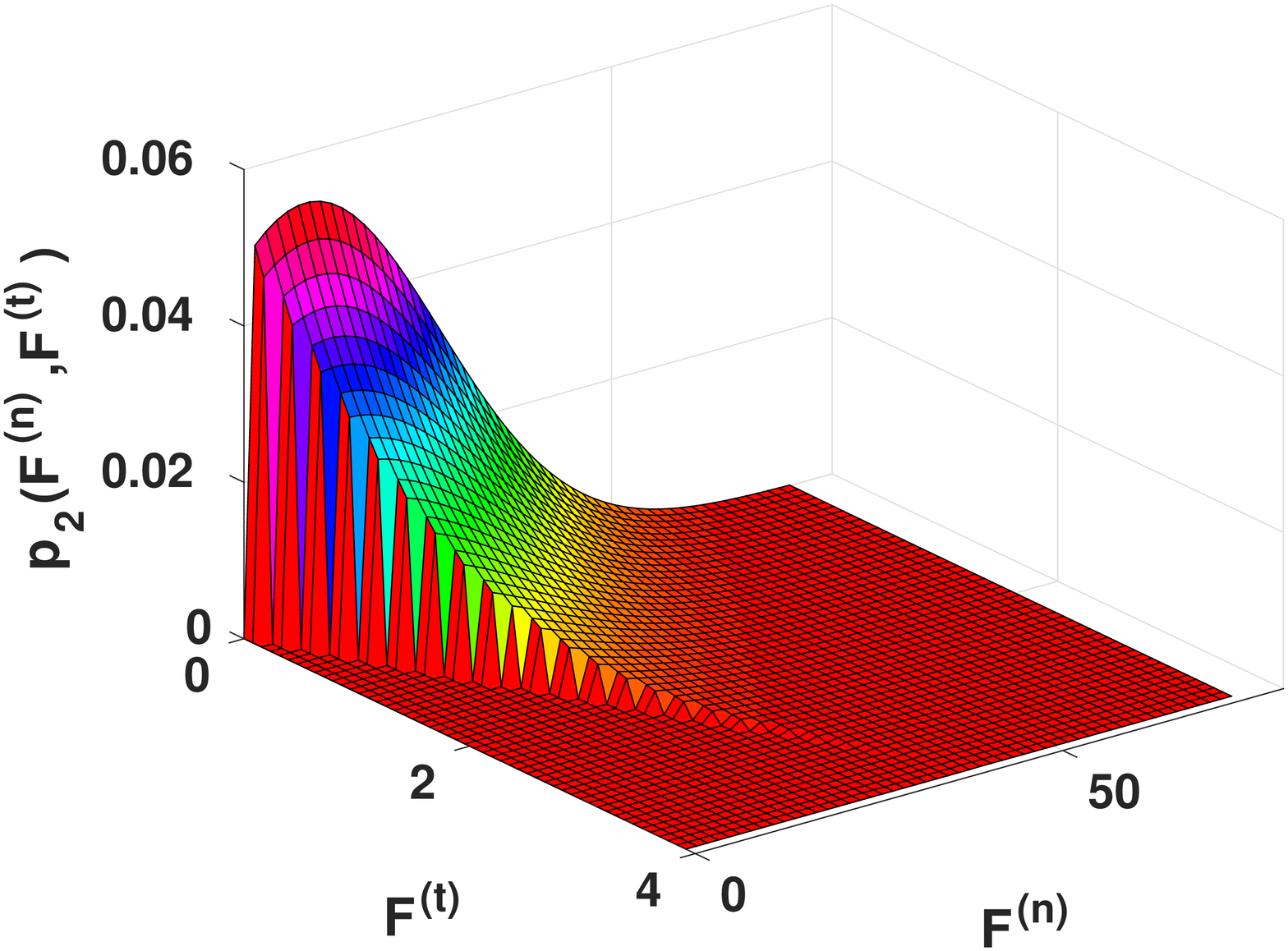}
\caption{Plot of the joint probability $P_2(F^{(n)},F^{(t)})$ versus $F^{(n)},F^{(t)}$ for  $\C P=83.5$ in the upper panel and $P_2(F^{(n)},F^{(t)})$  versus $F^{(n)},F^{(t)}$ for $\C P=20$ in the lower panel. Note  the early appearance of a singularity at small values of $F^{(n)},F^{(t)}$ at the lower pressures. Note also the very different scales required to plot $P_2(F^{(n)},F^{(t)})$ at high and low pressures. The sharp
drop in probability is due to the Coulomb constraint, cf. Eq.~(\ref{P2,a}).}
\label{jpplots}
\end{figure}
\subsection{Predictions}
\label{predict}

\begin{figure}
\includegraphics[scale=0.23]{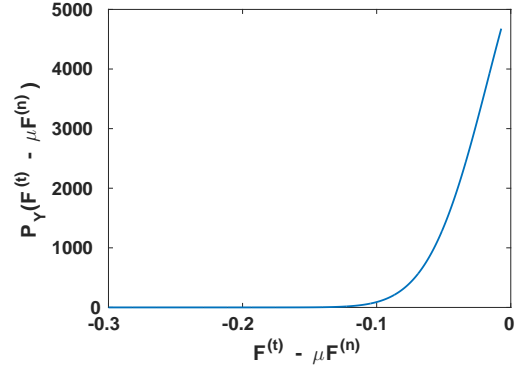}
\caption{Plot of the probability distribution for the variable $Y = F^{(t)} - \mu F^{(n)}$ given by $P_Y(Y;\C P=0.27)$   versus $Y$ for $\C P=0.27$ supporting the notion that a singularity appears at low pressures. }
\label{singplot}
\end{figure}

 We are now in a position to find the joint probability  $P_2(F^{(n)},F^{(t)})$ at different pressures from Eqs.~(\ref{P2,a}) and ~(\ref{Za}). Plotting the resulting join pdf's results in a very interesting
 observation: as the pressure reduces a singularity in the distribution starts to appear at low pressures (see Fig.~\ref{jpplots}). In addition the $F^{(n)},F^{(t)}$ axes contract (since the mean
 forces are proportional to the pressure). It appears that the joint probability diverges near $F^{(n)}=0$,$F^{(t)}=0$ as $P_2(0,0;\C P) \propto 1/{\C P}^2$.
The singular behaviour can be seen more directly in the variable $Y = F^{(t)} - \mu F^{(n)}$. In Fig.~\ref{singplot} we have plotted  $P_Y(F^{(t)}- \mu F^{(n)}; \C P)$ which can be found directly from the definition
\begin{eqnarray}
\label{Py}
 &&P_Y(F^{(t)} - \mu F^{(n)}; \C P)  \\ && =  \int_0^\infty \int_0^\infty dF^{(n)} dF^{(t)} P_2(F^{(n)},F^{(t)}; \C P) \delta (F^{(t)} - \mu F^{(n)})\ , \nonumber
 \end{eqnarray}
 for  $\C P=0.27$. It is clear that a singularity is growing at $Y=0$.

\subsection{validation and consequences}
\begin{figure}
\includegraphics[scale=0.70]{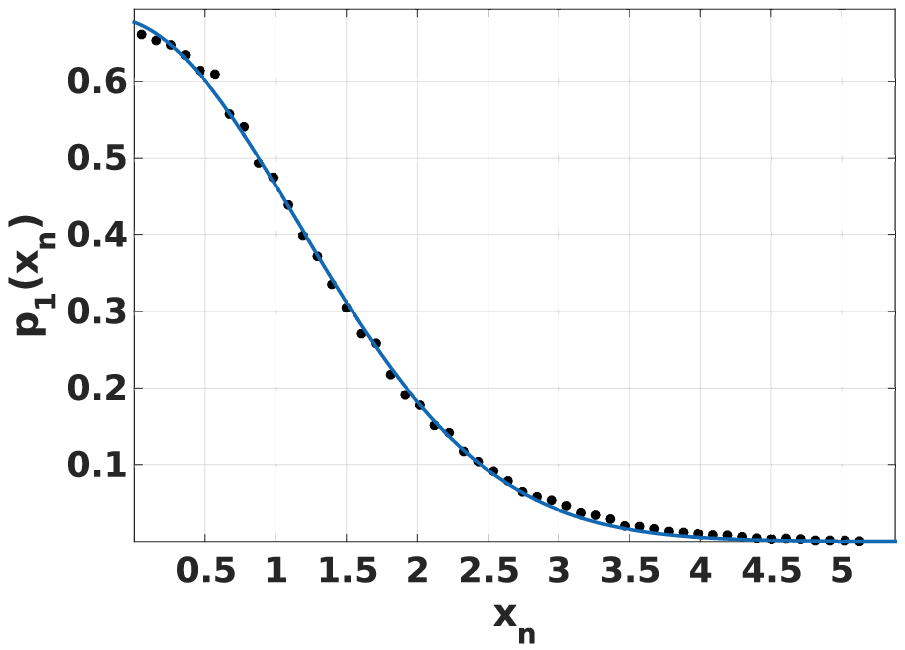}
\includegraphics[scale=0.70]{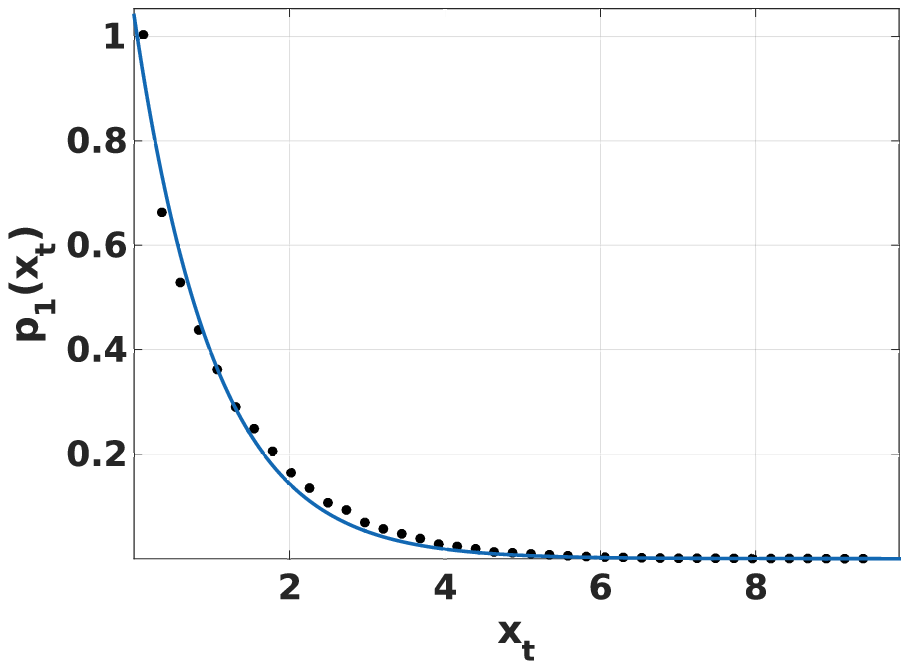}
\caption{The pdf's of the mean-normalized tangential forces as measured in the
simulations for the low pressure $\C P=0.27$.  In the upper
panel $\lambda_n=0.1$, $\lambda_{nn} = 0.28$. In the lower panel $\lambda_t=0.97$, $\lambda_{tt} = 0.01$. }
\label{low}
\end{figure}

\begin{figure}
\includegraphics[scale=0.22]{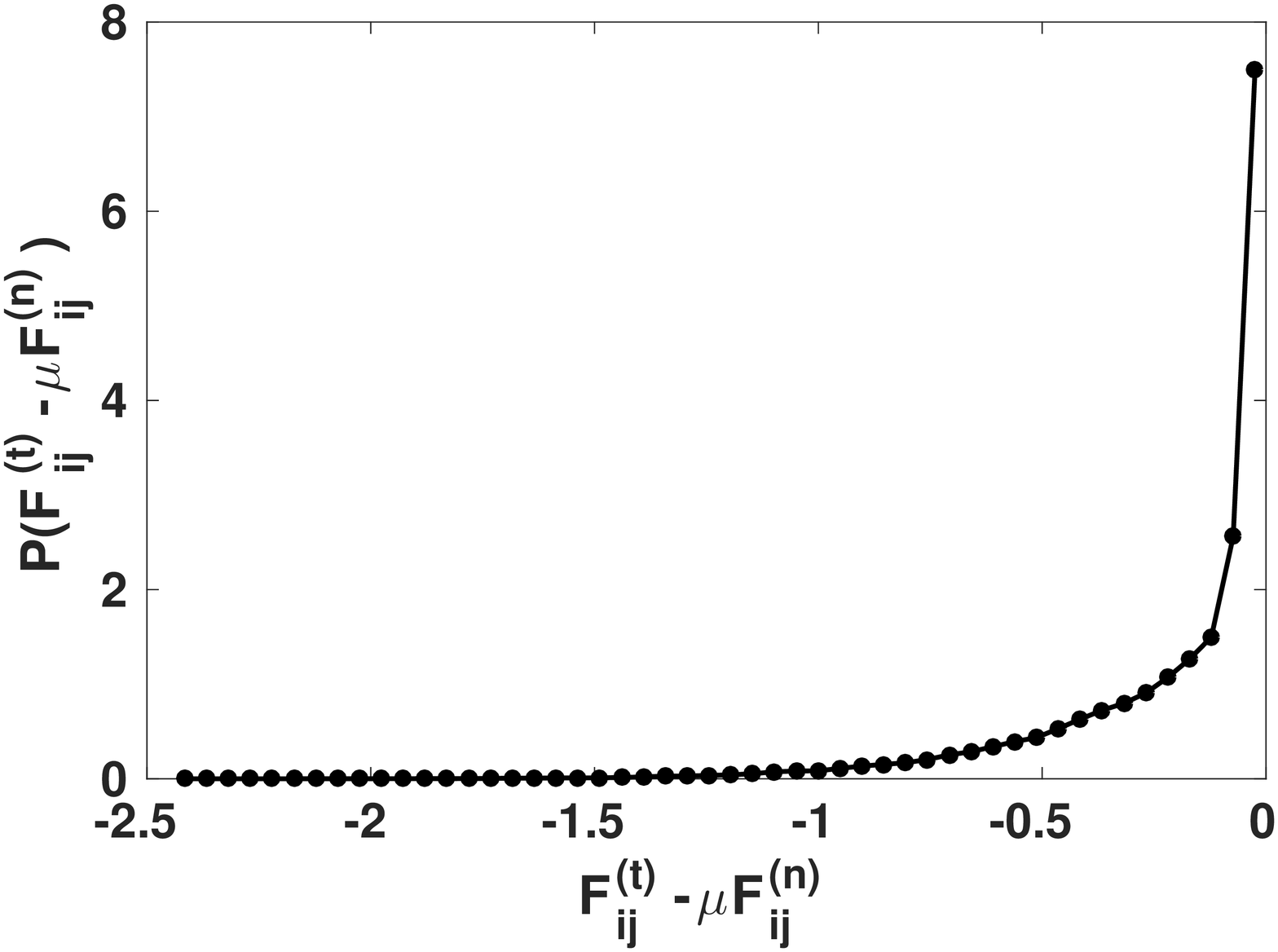}
\includegraphics[scale=0.22]{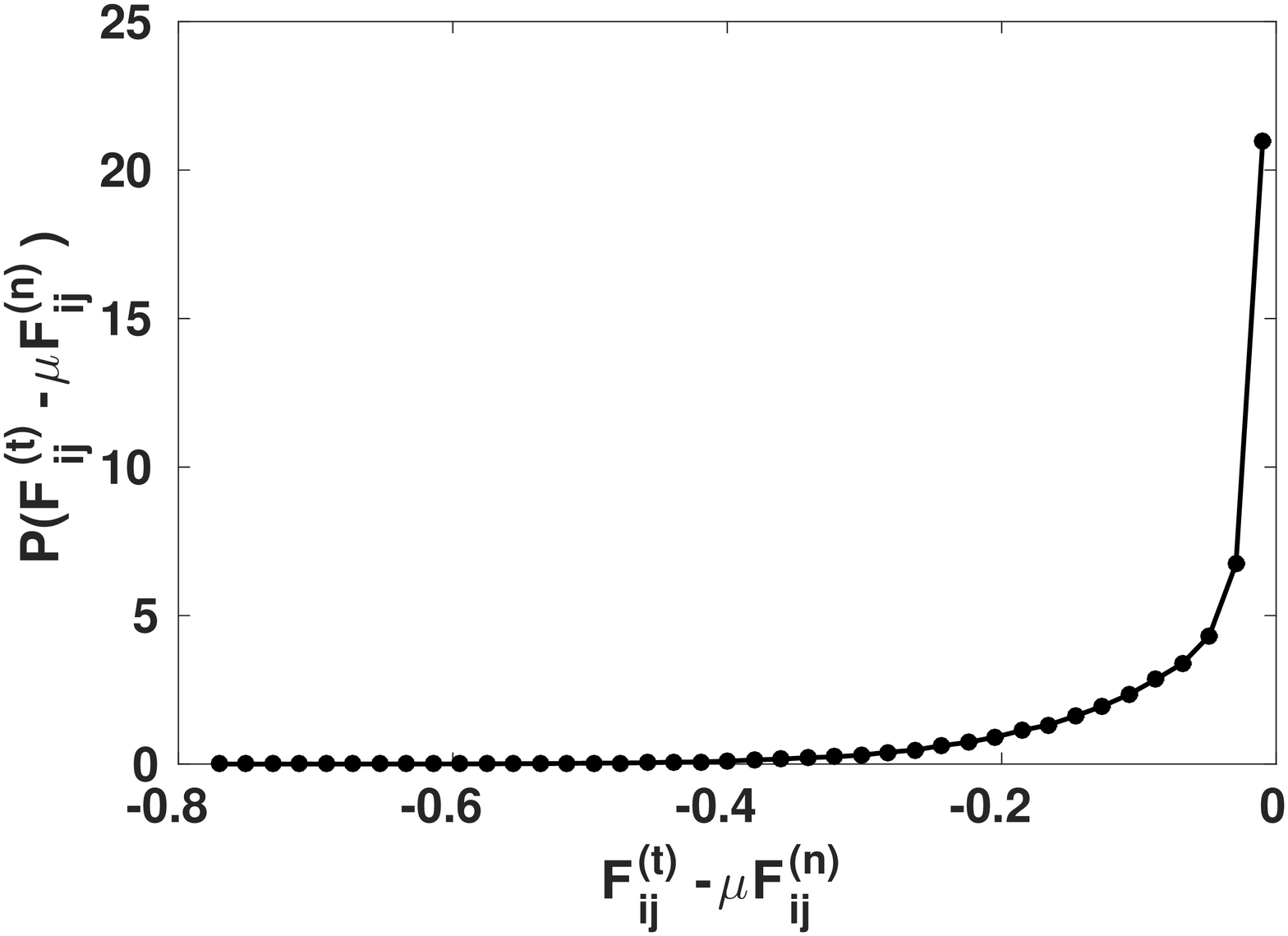}
\includegraphics[scale=0.22]{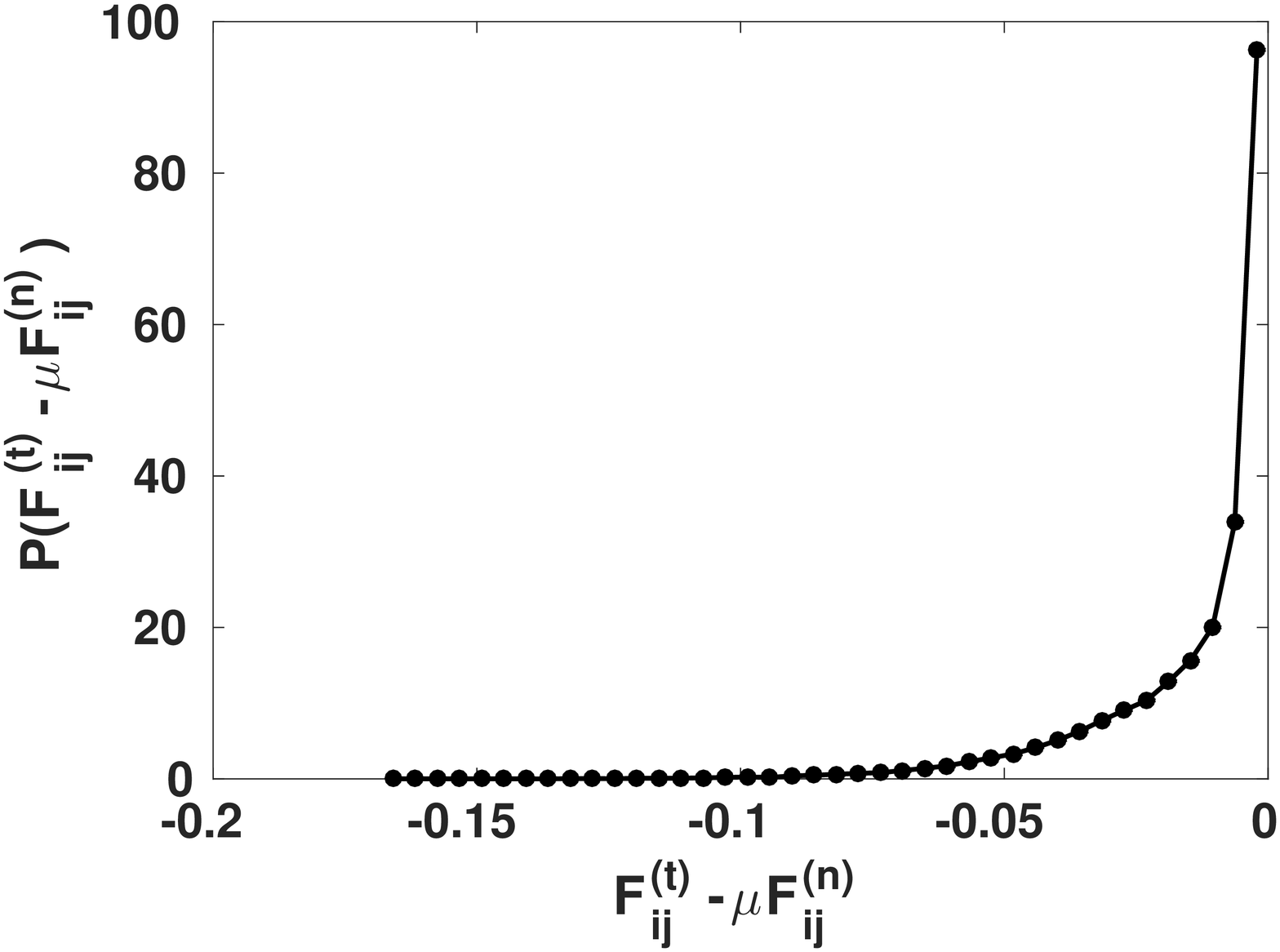}
\caption{The pdf $P( F^{(t)}_{ij}-\mu F^{(n)}_{ij})$ for varying pressures. Upper panel: $\C P=5$. Middle panel: $\C P=1$. Lower panel: $\C P=0.27$. Note the scale and the tendency for a singularity near zero when the pressure reduces.}
\label{probslip}
\end{figure}

The maximum entropy formalism predicts
an interesting and revealing aspect of the joint pdf's,  hidden
in their pressure dependence. To flush out this aspect we show in Fig.~\ref{low} the
pdf's $p_1(x_n)$ and $p_1(x_t)$ obtained in simulations for the low pressure $\C P=0.27$.
Besides the obvious remark that the maximum entropy forms fit the data very well also at this very low pressure, we can now see the systematics in $p_1(x_n)$ as a function of the pressure. Comparing the
upper panels of Figs.~\ref{comsim1}, \ref{comsim2} and \ref{low} we can see that the probability
to find {\em small} normal forces is increasing when the pressure decreases. We even lose at $\C P=0.27$ the maximum in $p_1(x_n)$ which is so prominent at higher pressures. But this means that when
the pressure reduces there can be a higher probability to bust the Coulomb conditions Eq.~(\ref{Coulomb}). Accordingly, we can expect that decreasing the pressure may result in large
frictional slip events. As argued in the last subsection the best way to examine this possibility is to use our numerics to compute the probability distribution function  $P( F^{(t)}_{ij}-\mu F^{(n)}_{ij})$. we display this function for varying pressures in Fig.~\ref{probslip}.
We observe the tendency of the pdf to exhibit a singularity near zero when the pressure
reduces. This is a strong indication that when we approach $\C P=0$ we should expect a giant
frictional slip event that is connected to the presence of an ``unjamming" singularity.

\begin{figure}
\includegraphics[scale=0.25]{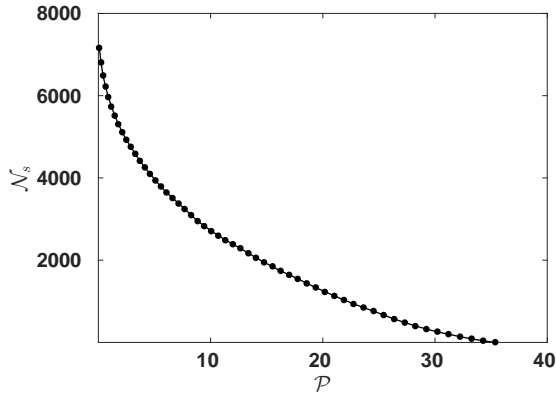}
\caption{The cumulative number of frictional slips ${\cal N}_s (N,\C P_{\rm max},\C P)$ as a function
of $P$ averaged over 10 independent decompression legs. The maximal pressure $\C P_{\rm max}$ averaged
 over these 10 legs is $\C P_{\rm max}=35.4$ and $N=4000$. Note the apparent
divergence of the cumulative number when the pressure reduces to zero.}
\label{result}
\end{figure}

To test this prediction we
focus now on a typical decompression protocol and ask how many frictional slip events ${\cal N}_s$ occur while we decompress from the maximal pressure to any given pressure $\C P$. In other words,
we measure
\begin{equation}
{\cal N}_s (N,\C P_{\rm max},\C P) \equiv \int_P^{\C P_{\rm max}} n(N,\C P) d\C P \,
\label{defNs}
\end{equation}
where $n(N,\C P) d\C P$ are the number of frictional slips that occur when decompressing
from $\C P+d\C P$ to $\C P$:
\begin{equation}
n(N,\C P) \equiv -\frac{d{\cal N}_s (N,\C P_{\rm max},\C P)}{d\C P} \ .
\end{equation}
The result of the measurement of ${\cal N}_s (N,\C P_{\rm max},\C P)$ as a function of $\C P$ is shown
in Fig.~\ref{result}. The simulation indicates an apparent divergence of the cumulative
number of slip events as the pressure reduces towards zero.

The quantitative theoretical understanding of the divergence of the cumulative slips will be
described in a later publication.

\section{Concluding remarks}
\label{conclusions}

In conclusion, we considered the pdf's for the magnitudes of normal and tangential forces in
frictional granular matter, focussing on the marginal and joint distribution $P_1(F^{(n)})$,
$P_1(F^{(t)})$ and $P_2(F^{(n)},F^{(t)})$. We showed that the maximum entropy formalism
provides a very adequate functional form for these pdf's in both experiment and simulations
at all the considered pressures. The fits were excellent when the pdf's exhibited maxima as
well as when maxima were absent. For the marginal pdf's two Lagrange multipliers were called
for, and five were necessary for the joint pdf's. Thus the mean and variance of the distributions
were also sufficient to provide the necessary Lagrange multipliers. In addition to reporting the
useful descriptive nature of the functional forms provide by the maximum entropy formalism, we
also presented their predictive usefulness. The formalism generated joint pdf's with increasing
singularity towards low pressure. This singularity indicated that giant frictional slips
are expected close to un-jamming. Simulations supported fully this prediction.

It would be useful in the future to examine the predictions of the maximum entropy formalism
in situation of different external strain like shear, oscillations etc. Taking into account
the quality of the fits presented above and the predictiveness of the resulting pdf's it seems
worthwhile to examine the range of applicability in both experiments and simulations.

\acknowledgments
This work has been supported in part by the ISF-Singapore program and the US-Israel BSF.
VSA, MMB, and the experiments were supported by the Collective Interactions Unit, OIST Graduate University. MMB gratefully acknowledges generous hosting by Prof. Surajit Sengupta at TCIS, TIFR while working on this paper.

\end{document}